\documentstyle [prl,twocolumn,aps] {revtex}

\begin{document}
\draft
\title{ Generalized Statistics and Solar Neutrinos }
\author{ G. Kaniadakis, A. Lavagno and P. Quarati 
\thanks{e-mail: Quarati@polito.it}}
\address{ Dipartimento di Fisica and INFM- Politecnico di Torino -
Corso Duca degli Abruzzi 24, 10129 Torino, Italy \\ 
Istituto Nazionale di Fisica Nucleare, Sezioni di Cagliari e di Torino}
\maketitle
\begin {abstract} {\bf Abstract:} 
The generalized Tsallis statistics produces a distribution function 
appropriate to describe the interior solar plasma, thought as a 
stellar polytrope, showing a tail depleted respect to the 
Maxwell-Boltzmann distribution and reduces to zero at 
energies  greater than about $20 \, k_{_B} T$. The Tsallis statistics can 
theoretically support the 
distribution suggested in the past by Clayton and collaborators, 
which shows also a 
depleted tail, to explain the solar neutrino counting rate.
\end {abstract}

\noindent

\vspace{1cm}

By imposing a variational principle one can conveniently generalize both 
Boltzmann-Gibbs statistics and standard thermodynamics \cite{tsa,cura}. 
The lack of 
adequacy of the Boltzmann entropy is related to the breakdown of 
Boltzmann-Gibbs 
statistics for systems with long-range interactions \cite{pla,nob} such as, 
among 
others, the interior solar plasma, where the particles are exposed to 
many-body collisions and the energy available to a particular pair of 
particles is 
not defined \cite{cla2}.

Even for a non perfect plasma, the ideal gas approximation 
is usually considered a good one. 
The Maxwell-Boltzmann (MB) distribution is believed to be highly correct 
and the many-body 
physics involved does not determine sensible deviations from the standard 
statistics 
used \cite{kip}.
However, non-Maxewellian (flattened) distributions in plasmas heated by 
inverse 
bremsstrahlung (collisional) absorption of sufficiently strong 
electromagnetic fields 
have been predicted and recently measured \cite{mat,erd,zito}; 
in $D$ - $^3 He$ 
fusion plasma experiments, a nonthermal component of the ion distribution 
and the possible 
consequences on the nuclear rates have been investigated \cite{fis,lap}.

Among the distributions that deviate from the MB one, we pose our attention 
on distributions coming from generalized statistics.
We give a very brief outline of the Tsallis generalization of 
thermodynamics and statistical physics \cite{tsa,cura}, 
suitable for describing systems with 
long-range interactions \cite{pla,nob}.\\
If we have a system with $W$ microscopic states, each with probability 
$f_i \ge 0$ 
normalized as
\begin{equation}
\sum_{i=1}^{W} f_i=1\ \ ,
\end{equation}
then the entropy is given by
\begin{equation}
S_q=\frac{k}{q-1} \sum_{i=1}^W f_i (1-f_i^{q-1})\ \ ,
\end{equation}
where $k$  and $q$  are constants. In the limit as  $q$  approaches unity, 
the well 
known expression $S_{1}=-k \sum_i f_i \log f_i$ is recovered and we can fix 
$k=k_{_B}$ (the Boltzmann constant), i.e. the Tsallis statistics reduces 
to the standard one as $q\rightarrow 1$.

We have shown a generalization of the Tsallis statistics by using a kinetic 
approach, 
based on the Fokker-Planck equation, and we have given to the Tsallis 
parameter  $q$  
the meaning of a measure of the deviation from constant behavior 
of the diffusion coefficient $D(v)$, 
a quadratic function of the velocity $v$ \cite{ka1,ka2,ka3}. 
The Tsallis generalized  statistics is actually widely used in many different 
physical 
problems (we send the reader to Ref.s \cite{bar,levy} where many references 
on the different applications are quoted).

One of the first problems where Tsallis statistics has been applied is that 
of stellar 
polytropes \cite{pla}. Nobre and Tsallis \cite{nob} and other authors 
have recently 
discussed the physical needs for departure from Boltzmann-Gibbs statistical 
mechanics and thermodynamics for gravitational like systems \cite{bog} and 
anomalous diffusion \cite{levy,zan}
(this last subject could be of great interest in the study of solar core, 
due to the 
well known problem of the diffusion of light elements like $Li$ and $Be$ 
\cite{ric}). 

Plastino and Plastino \cite{pla} have sought help from Tsallis entropy to find 
sensible distribution functions for stellar polytropes while that of MB gives 
unphysical distribution functions.\\
They found a range of variability of $q$ from a relation between 
the polytropic index $n$ and $q$ deduced 
comparing two different but equivalent expressions of the 
distribution. 
We obtain a different relation between $n$ and $q$ because 
we impose a special constraint on the solar gravitational potential.\\
The internal structure of the sun can be considered polytropic 
with index $n$ ranging 
between the value 3/2 ($\gamma$=5/3) and 5 ($\gamma$=6/5)\cite{kip}
($\gamma$ is the adiabatic parameter).
In the interior regions where hydrogen ionization is changing rapidly, 
$\gamma$ is 
taken to be very close to unity ($n=\infty$, $q=1$), therefore in these 
circumstances the MB 
distribution holds.

We want to apply generalized statistics to derive a distribution function 
for the interior solar plasma, relevant in 
calculating the nuclear fusion reaction 
rates responsible for the neutrino flux emitted by the sun \cite{bah2,gal}.
The following arguments support the program of this work.

It is well known that one of the main problems with solar physics 
is related to  
the detected neutrinos arriving from the sun. All the different 
experiments have 
confirmed a deficit in the flux relative to the predictions of standard 
theory of 
nuclear physics \cite{ric,gal}. In particular, the neutrino flux from 
the reactions 
involving $^7 Be$ and $^8 B$, mainly due to collisions at energies higher 
than the 
effective energy of $pp$ reactions (4.58 $k_{_B} T$), is much lower 
than that predicted by the standard 
models, which uses MB distributions \cite{ric}.

In the recent past we have tried to contribute to the solution of the solar 
neutrino problem 
deriving, as steady state solutions of the Fokker-Planck equation, statistical 
distributions that differ from the MB distribution for a depleted tail 
at high energies \cite{ka1,ka3}.
The distribution $f$ is also the solution of a Boltzmann equation without 
collisions 
(at equilibrium the distributions obtained with and without the collision term 
coincide).\\
Non-Maxwellian distribution can give 
thermonuclear reaction rates $r=<\sigma v>$ 
smaller than the standard ones \cite{lap}, allowing a reduced flux of 
neutrinos from the sun interior, particularly at energies above few $k_{_B} T$.

Very recently we have shown that the Tsallis distribution is a particular case 
of the 
family of distributions we derived by means of the Fokker-Planck 
equation kinetic 
approach \cite{ka2,ka3}. 
A well defined statistics can be derived for any particular pair of 
values (M,N) 
indicating the degree of the polynomials, in the velocity variable, used 
to describe 
the drift $J$ and diffusion $D$ coefficients. The Tsallis classical and quantum 
statistics are related to the pair (0,1) (in this paper we do not derive 
the drift and 
diffusion coefficients of the solar core, this will be investigated 
elsewhere, rather 
we use straight way the generalized Tsallis statistics).

In addition to this, let us recall that Clayton and collaborators 
\cite{cla2,cla3} 
suggested, on phenomenological grounds, that neutrino counting rate 
will be much 
reduced if the high energy tail of the MB distribution of relative 
energies is depleted,
 the depletion being described by the factor $\exp \{-\delta \, 
(E/k_{_B}T)^2\}$ with 
a suggested value $\delta \approx 0.01$.\\
The distribution that we have derived in Ref.\cite{ka1,ka2}, of the 
Tsallis type, can 
take into account this behavior, certainly due to the long-range 
gravitational 
interaction. The Tsallis parameter $q$ can be related to the Clayton parameter 
$\delta=(1-q)/2$; our constraints impose a value 
of $\delta$ slightly different from the value argued by Clayton. 
We obtain that the value of $\delta$ must satisfy the range of 
variability $0.02 \le \delta \le 0.05$.

To show the analogies of our approach with the distribution suggested 
by Clayton, 
we introduce, by means of an expansion, an approximated 
expression of 
the Tsallis distribution 
which is correct in the range of energies of interest here. \\
The Tsallis distribution function reduces to zero at 
$E \approx (10\div 25) \, k_{_B}T$ depending on the value of $\delta$; 
therefore, 
high energy collisions are greatly reduced or absent and the neutrino flux 
from $^7 Be$ and $^8 B$ reactions is smaller than foreseen by standard 
theories, the 
importance of the neutrino flux from $pp$ reactions is consequently increased. 

In this work we show that the suggestion of Clayton and collaborators is well 
founded on theoretical grounds (long-range gravitational interaction), it is 
equivalent to using the distribution we have derived elsewhere 
\cite{ka1,ka2,ka3} and that 
this distribution is well motivated within the Tsallis statistics.
The neutrino counting rate, compared to the standard predictions, 
can be explained 
using the above prescriptions. 
A complete treatment of this subject can be carried on only 
within a complete solar model code \cite{ich,cas,tur}, it will be done and 
reported elsewhere.

\vspace{1cm}

Following Tsallis \cite{tsa,cura} Plastino and Plastino \cite{pla} and 
Boghosian \cite{bog}, the distribution 
function $f$ of 
stellar polytropes is 
\begin{equation}
f\propto \left [ 1+(q-1)(\alpha+\beta\epsilon) \right ] ^{1/(1-q)} \ \ ,
\end{equation}
where $q$ is the Tsallis parameter and $\epsilon=E+\Phi(r)$, 
where $E$ is the c.m. 
kinetic energy, $\Phi(r)$ is the gravitational potential, 
$\alpha=-\beta\mu$, where $\mu$ is the chemical potential, 
$\beta=1/(k_{_B}T)$. The distribution (3) becomes the MB distribution 
$\exp[-(\alpha+\beta\epsilon)]$ as $q$ goes to one.
Let us define the relative gravitational potential $\Psi$ 
\begin{equation}
\Psi=-\Phi+\Phi_0 \ \ ,
\end{equation}
where $\Phi_0$ is a constant to be chosen such that $\Psi$ vanishes at 
the edge of the system \cite{kip}. The following relation 
between the relative potential $\Psi$ and the density $\rho$ holds
\begin{equation}
\rho^{\gamma-1}=\frac{\gamma-1}{{\cal K}\gamma} \Psi \ \ ,
\end{equation}
where ${\cal K}={\cal P}/\rho^{\gamma}$, ${\cal P}$ is the pressure 
and $\gamma$ is the adiabatic parameter.\\
The polytropic index $n$ is defined by
\begin{equation}
\gamma=1+\frac{1}{n} \ \ .
\end{equation}
We introduce the relative energy ${\cal E}=\Psi-E$, 
the distribution $f$ can be written as
\begin{equation}
f\propto \left [ 1+(q-1)(\alpha+\beta\Phi_0)-(q-1)\beta{\cal E} 
\right ]^{1/(1-q)} \ \ .
\end{equation}
The quantity $\alpha$ can be chosen, without loosing generality, 
in such a way that \cite{bog}
\begin{equation}
1+(q-1)(\alpha+\beta\Phi_0)=0 \ \ .
\end{equation}

We want to compare the Tsallis statistics $f$ to the distribution 
introduced by Clayton et al. 
\cite{cla3} 
which is equal to the MB distribution $C \, \exp(-\beta E)$ times a 
correction factor
\begin{equation}
f=C \, e^{-\beta E} \, e^{\varphi(E)}=e^{\beta_0-\beta_1 E-\beta_2 E^2} \ \ ,
\end{equation}
where we set $\beta_1=\beta+B$ ($B$ is a constant).\\
We observe that the Tsallis distribution can be written in a classical form 
$f\propto e^{-\hat\epsilon}$ where
\begin{equation}
-\hat\epsilon=\frac{1}{1-q} \left\{\log [1+(q-1)(\alpha+\beta\epsilon)]
\right\} \ \ .
\end{equation}
The function $\hat\epsilon$ may be interpreted as a 
generalized energy that takes into account many body 
collective interactions \cite{lav}.\\
The logarithmic term in Eq.(10) can be 
expanded in powers of $(q-1)(\alpha+\beta\epsilon)$  with the conditions 
$\mid (q-1)(\alpha+\beta\epsilon)\mid \le 1$ and 
$(q-1)(\alpha+\beta\epsilon) \ne -1$.\\
We impose that 
\begin{equation}
\alpha+\beta\Phi=0 \ \ ,
\end{equation}
to allow that $\beta_1=\beta$, as required 
for physical reasons in the approximated distribution by Clayton 
\cite{cla2,cla3}.\\
Infact, $B$ does not have any effect and can be ignored because we maintain 
the solar 
luminosity at its known value and the increase of the central temperature 
$T$ 
to counteract the effect of $B$ on the power of the sun raises the $^8 B$ 
neutrino flux back to the value it had at $B=0$.\\
The special condition (11) or $\alpha=\beta(\Psi-\Phi_0)$, which is 
a constraint characteristic of the solar core, shows that $\alpha$ is a 
function of $\Psi$ 
or $({\cal E}+E)$, therefore the comparison between two different 
but 
equivalent expressions of $f$ done to derive a relation between 
$n$ and $q$ in Ref.s\cite{pla,bog} is not allowed in this case. 
Finally, we find that the parameters $\beta_0$, $\beta_1$, $\beta_2$ are: 
$\beta_0=\log C$, $\beta_1=\beta$ and $\beta_2=(1-q) \beta^2 
/2=\delta\beta^2$.\\ 
The above constraints (8) and (11) imply
\begin{equation}
1+(q-1)\beta\Psi=0 \ \ ,
\end{equation}
and from Eq.s (5), (6) and (12) we obtain
\begin{equation}
q=1-\frac{\tau}{n+1}  \ \  {\rm or} \ \ \  \delta \, (n+1)=\frac{\tau}{2} \ \ ,
\end{equation}
with $\tau=k_{_B} T \, \rho/\cal P$.
This relation, which links $n$ and $q$, differs from the relation 
given by Plastino and Plastino \cite{pla} or from the one given by 
Boghosian \cite{bog} because of the constraint (11) valid for the solar core.


In Ref.\cite{kip} it is shown that, if we select $n=3$, the following values of 
$\rho$, $\cal P$  and $T$ of the solar core can be derived: the proton density 
$\rho=53.47$ gr/cm$^3=0.32 \, 10^{-13}$ protons/fm$^3$, the pressure 
${\cal P}=0.77$ $10^{-16}$ MeV/fm$^{-3}$ and the temperature 
$k_{_B}T=1.034$ $10^{-3}$ MeV which is slightly lower than the temperature 
fixed by other models ($1.29$ $10^{-3}$ MeV). These last two figures are 
determined supposing a particular central composition; changing the 
composition by increasing the $He$ concentration a greater temperature 
can be reached. \\
With this selection of values we obtain $\tau=0.43$ and $\delta=0.05$.

In the solar core the value of the different physical quantities of interest in 
this work have 
been reported by Ichimaru \cite{ich} to be: $k_{_B}T=1.29 \,\, 10^{-3}$ MeV, 
$\beta=0.77 \,\, 
10^3$ MeV$^{-1}$, $\rho=56.2$ gr/cm$^{3}$, which is the 36\% of 156 gr/cm$^3$, 
for the free protons and ${\cal P}=2.12 \, \, 10^{-16}$ MeV/fm$^3$.\\
By inserting these values into Eq.(13) we obtain (with $n=3$) $\tau=0.2$ and 
$\delta (n+1)=0.1$; then we can fix the central composition to maintain the 
magnitude of 
$\tau$  to be $0.2$ and select different couples of values of $n$ and $\delta$, 
e.g.: $n=3/2$ and $\delta=0.04$, $n=3$ and $\delta=0.025$, $n=5$ and
$\delta=0.017$. 
In conclusion, we may expect that the appropriate value of $\delta$ be in the
 range between $0.02$ and $0.05$. The value proposed by Clayton 
($\delta=0.01$) gives (when $\tau=0.2$) 
$n=9$ which is unphysical, because greater than $5$.


The approximated distribution function with $\delta=0.02$ is
\begin{equation}
f_{approx}=C^{'} \, e^{-0.774 \, E_{\rm keV}-0.012 \, E^2_{\rm keV}} \ \ ,
\end{equation}
where $C^{'}$ is the normalization constant.
It is easy to verify that the correct Tsallis distribution reduces to zero at 
$E=k_{_B}T/(2\delta)$ forbidding the existence of ions with energies 
greater than 
$10\div 25$ times the sun temperature.
We wish to recall that the Eq.(14) represents a Druyvenstein distribution 
\cite{dru}.\\ 
All quantities relevant in solar physics and solar neutrino emission can 
be expressed and evaluated as functions of the parameter $\delta$.
We report here, for instance, the thermonuclear reaction rate 
corrected respect to the MB 
rate $r_{_{MB}}$ by the depletion factor and calculated in Ref.\cite{cla3}
\begin{equation}
r=r_{_{MB}}\, (1+\frac{15}{4} \delta-\frac{7}{3}\delta 
\frac{E_0}{k_{_B}T}+\cdots)\, e^{-\Delta} \ \ ,
\end{equation}
where $E_0$ is the most effective energy ($E_0=4.5$ $k_{_B}T$ 
for $pp$ reactions) 
and $\Delta$ is a function of $E_0$, $ k_{_B}T $ and $\delta$.

The neutrino counting can be reduced sensibly above few $ k_{_B}T$ with  some 
changes in the solar model parameters. These can be compensated by small 
changes in the initial $He$ concentration. Most of this reduction has 
come at
the expenses of the $^7 Be$ and $^8 B$ neutrino fluxes. Counting rates 
within the 
solar model used by Clayton and collaborators can be extrapolated from 
the curves 
reported in Fig.2 of their work \cite{cla3}.
A value of $\delta$ different from zero makes a star more luminous and 
reduces the 
rate of energy production at a given temperature. The solar core contracts 
to higher 
temperature. As shown by Clayton et al. the increase of temperature at given 
solar luminosity does not increase the neutrino fluxes that decrease 
with $\delta$.

We do not discuss further, in this work, the measured results and the 
predictions of the neutrino fluxes; we leave complete and more definitive 
discussion 
after the proposed distribution will be tested within the available solar 
models \cite{bah2,cas,tur,cla4}.
Of course, the results reported in this work depend on the values of 
the parameters
of the solar core one takes as input.
We can expect that the trend of definitive results will be on the line of 
the present 
description. We hope that the content of this work could be useful to the 
operating and proposed solar neutrino and underground nuclear 
astrophysics experiments \cite{arp1,arp2}.

\vspace{1cm}

We thank B. Boghosian, D. Clayton, A. Erdas, G. Fiorentini, M. Lissia and 
C. Tsallis for critical reading of the manuscript, comments and discussions.

\end{document}